\documentclass[%
 reprint,
 amsmath,amssymb,
 aps,
]{revtex4-2}

\usepackage{graphicx}
\usepackage{dcolumn}
\usepackage{bm}
\usepackage{subfigure}
\usepackage[dvipsnames]{xcolor}
\usepackage{amssymb,mathtools}
\usepackage{bigints}

\begin{document}

\preprint{APS/123-QED}

\title{Optimal Construction of Two-Qubit Gates using the Symmetries of B Gate Equivalence Class}

\author{M. Karthick Selvan}
\email{karthick.selvan@yahoo.com}%

\author{S. Balakrishnan}%
 \email{physicsbalki@gmail.com}
\affiliation{Department of Physics, School of Advanced Sciences, Vellore Institute of Technology, Vellore - 632014, Tamilnadu, India.}%



\begin{abstract}
Two applications of gates from the B gate equivalence class can generate all two-qubit gates. This local equivalence class is invariant under the mirror (multiplication with the SWAP gate) operation, inverse (Hermitian conjugate) operation, and the combined inverse and mirror operations. The last two symmetries are associated with the ability of a two-qubit gate to generate the two-qubit local gates and the SWAP gate in two applications. No single local equivalence class of two-qubit gates, except the B gate equivalence class, has these two symmetries. Only the planar regions of the Weyl chamber, describing the mirror operation, contain the local equivalence classes with either one of the two symmetries. We show that there exist one-parameter families of local equivalence classes on these planes, with and without the B gate equivalence class, such that each of them can be used to construct a parameterized universal two-qubit quantum circuit that involves only two nonlocal two-qubit gates. We also discuss the implementation of the gates from a few families of local equivalence classes on superconducting quantum computers for optimal generation of all two-qubit gates. We provide upper bounds on the number of two-qubit gates required to generate an arbitrary $n$-qubit gate for two families, each of which is conjectured to generate all two-qubit gates in two applications. We show that there exists a positive correlation between the area of the convex hull of the squared eigenvalues of the nonlocal part of a parameterized two-qubit gate and the fractional volume of the Weyl chamber covered in two applications of the parameterized two-qubit gate for two families of local equivalence classes. 
\end{abstract}

\maketitle

\section{Introduction}
We study the optimal construction of two-qubit gates using one-parameter families of two-qubit gates. By optimal construction, we mean the generation of a given two-qubit gate by two applications of a capable gate from the family that can be implemented with better fidelity. The motivation of this study is the following. Noisy intermediate-scale quantum (NISQ) processors use a fixed basis gate set that involves an entangling two-qubit gate and a set of single-qubit gates for performing quantum computation. The performance of the basis set majorly depends on the choice of the entangling two-qubit gate. Any entangling two-qubit gate can be used as a basis gate for doing quantum computation~\cite{Deutsch1995,Lloyd1995}. The choice of an entangling two-qubit gate as a basis gate depends on the type of interaction natively realizable on the given quantum processor and the ability of the entangling two-qubit gate to generate all two-qubit gates. In a given quantum processor, a family of entangling two-qubit gates can be generated by the native interaction. Among them, the entangling two-qubit gate, which can generate all two-qubit gates with the fewest applications and can be implemented with higher fidelity, is preferred as a basis gate. 

In the quantum processors, where the native interaction is locally equivalent to the $XX$ interaction~\cite{Rigetti2010,Chow2011}, though gates belonging to all controlled unitary equivalence classes can readily be implemented, a gate from the CNOT equivalence class is used as a basis gate, as they can generate all two-qubit gates in three applications~\cite{Vidal2004,Coffey2008}. The $XX+YY$ interaction can be used to generate both the $\sqrt{\text{iSWAP}}$ and iSWAP gates. Both of them can generate all two-qubit gates in three applications. However, in the quantum processor, where the $XX+YY$ interaction is native, the $\sqrt{\text{iSWAP}}$ gate was shown to be a better basis gate than the iSWAP gate, as the former can be implemented in a shorter duration than the latter~\cite {Huang2023}.

The advantage of a fixed gate set is that it simplifies the calibration process. Its drawback is that it uses only a single entangling gate to implement any unitary operation. There exist entangling two-qubit gates that can generate certain two-qubit gates in the same number of applications as the entangling basis gate, but can be implemented with better fidelity by the native interaction. Inclusion of such entangling gates in the basis set was shown to improve the fidelity of implementation of quantum circuits~\cite{Satoh2022}. The advantage of adding fractional controlled $X$ $(CX)$ gates to the basis set along with the $CX$ gate was studied analytically~\cite{Peterson2022}. In recent times, many works have been done on realizing a continuous basis gate set. A pulse-level control technique was proposed to implement continuously tunable cross-resonance gate~\cite{Rigetti2010,Chow2011} in superconducting processors, and its advantage was demonstrated~\cite{Earnest2021,Stenger2021,Chen2022,Sugawara2025}. Implementation of gates from the fermionic simulation gate family has also been reported~\cite{Foxen2020}. Ion-trap quantum processors also support continuous basis gate set~\cite{Debnath2016,Maslov2017}. Protocols for implementing any two-qubit gate between two superconducting qubits using a tunable coupler were proposed and demonstrated~\cite{Chen2024,Chen2025}. Techniques to implement any two-qubit gate between two superconducting qubits with fixed coupling were also demonstrated~\cite{Wei2024}. However, calibration of experimental parameters for all two-qubit gates is a difficult task. 

The optimal strategy is to use the family of gates that can be implemented by the native interaction of the quantum processor as a continuous basis gate set. The most capable gates from families implementable using $XX$ and $XX+YY$ interactions must be used three times to generate most two-qubit gates. Hence, it is necessary to identify families of two-qubit gates that can generate all two-qubit gates in two applications, and the schemes to implement them on different quantum processors. 

In the Weyl chamber~\cite{Zhang2003}, the geometry of local equivalence classes of two-qubit gates, the operations such as Hermitian conjugate (inverse operation) and multiplication with the SWAP gate (mirror operation), can be described as reflections with respect to certain planes passing through the Weyl chamber~\cite{Selvan2023}. For a two-qubit gate to generate two-qubit local gates and the SWAP gate in two applications, the point representing its local equivalence class, in the Weyl chamber, should be invariant under inverse and the combined inverse and mirror operations, respectively. The point representing the B gate local equivalence class is the only point common to all the reflecting planes and hence invariant under inverse, mirror operations, and their combination~\cite{Chen2024,Selvan2023}. The local equivalence classes having one of these two symmetries exist on the planes describing these symmetries. We conjecture that one-parameter two-qubit gate families can be formed from the local equivalence classes represented by these planes such that each of them can optimally generate all two-qubit gates. We provide examples and prove their ability to generate all two-qubit gates using the method described in Ref.~\cite{Peterson2020}. We discuss existing schemes for implementing the gates from these families of local equivalence classes on superconducting quantum processors. We obtain the upper bound on the number of two-qubit gates required to generate an arbitrary $n$-qubit gate for two specific families containing fermionic simulation gates. These upper bounds are valid provided all two-qubit gates can be generated by two applications of gates from the families. We also point out a positive correlation between the areas of the squared eigenvalues of the nonlocal part of two-qubit gates and the fractional volume of the Weyl chamber in two applications for specific families.


\section{Background}

\subsection{Local Equivalence Classes of Two-Qubit Gates}

Any two-qubit gate, $U \in \text{U}(4)$, can be written in the following form~\cite{Zhang2003}. 
\begin{equation}\label{eq1}
U = e^{i \alpha}\left[ k_1 \otimes k_2 \right] e^{\left\{ i H(c_1, c_2, c_3)/2 \right\}} \left[k_3 \otimes k_4 \right],
\end{equation}
where $e^{i \alpha}$ is the global phase factor, $k_i'\text{s} \in \text{SU}(2)$ $(i=1,2,3,4)$ are single-qubit gates, and the exponential term in the middle with 
\begin{equation}\label{eq2}
H(c_1, c_2, c_3) = c_1 (\sigma_x \otimes \sigma_x) + c_2 (\sigma_y \otimes \sigma_y) + c_3 (\sigma_z \otimes \sigma_z)
\end{equation}
is the nonlocal part of $U$, and is an element of $\text{SU}(4)/\text{SU}(2) \otimes \text{SU}(2)$. The triples $(c_1, c_2, c_3)$ are called Cartan coordinates; they describe the nonlocal characteristics of $U$. 

The eigendecomposition of $H$ can be written as 
\begin{equation}\label{eq3}
H(c_1, c_2, c_3) = \sum_{j=1}^4 h_j \vert \Psi_j \rangle \langle \Psi_j \vert,
\end{equation}
where the eigenvalue and the corresponding eigenvector pairs $\{h_j, \vert \Psi_j \rangle \}'\text{s}$ are as follows. 
\begin{equation}\label{eq4}
h_1 = c_1 - c_2 + c_3,~ \vert \Psi_1 \rangle = \dfrac{\vert 00 \rangle + \vert 11 \rangle}{\sqrt{2}},
\end{equation}
\begin{equation}\label{eq5}
h_2 = c_1 + c_2 - c_3,~ \vert \Psi_2 \rangle = \dfrac{i \vert 01 \rangle + i\vert 10 \rangle}{\sqrt{2}},
\end{equation}
\begin{equation}\label{eq6}
h_3 = - c_1 - c_2 - c_3,~ \vert \Psi_3 \rangle = \dfrac{ \vert 01 \rangle - \vert 10 \rangle}{\sqrt{2}},
\end{equation}
and
\begin{equation}\label{eq7}
h_4 = - c_1 + c_2 + c_3,~ \vert \Psi_4 \rangle = \dfrac{i \vert 00 \rangle - i\vert 11 \rangle}{\sqrt{2}}. 
\end{equation}

The eigenbasis of $H$ is also the eigenbasis of $U_d$. The local invariants of two-qubit gates are described using the eigenbasis of $U_d$ as follows~\cite{Zhang2003,Makhlin2002}. 
\begin{equation}\label{eq8}
G_1 = \dfrac{tr^2[m(U)]}{16 \det(U)},
\end{equation}
and
\begin{equation}\label{eq9}
G_2 = \dfrac{tr^2[m(U)] - tr[m^2(U)]}{4 \det(U)},
\end{equation}
where $m(U) = \left(Q^\dagger U Q \right)^T \left(Q^\dagger U Q \right)$ with the matrix $Q$ formed by the eigenbasis of $U_d$ as shown below.
\begin{equation}\label{eq10}
Q = \dfrac{1}{\sqrt{2}} \begin{bmatrix}
1 & 0 & 0 & i \\ 0 & i & 1 & 0 \\ 0 & i & -1 & 0 \\ 1 & 0 & 0 & -i 
\end{bmatrix}
\end{equation}

The local invariants $G_1$ and $G_2$ are complex-valued and real-valued functions, respectively. Two-qubit gates having the same values of $G_1$ and $G_2$ have the same nonlocal part and form a local equivalence class. Geometrically, each local equivalence class is represented as a point of a tetrahedron in the space spanned by the Cartan coordinates, as shown in FIG.~\ref{fig1}. This tetrahedron is referred to as the Weyl chamber in the positive Cartan coordinate system, where the Cartan coordinates satisfy the following conditions. 

\begin{equation*}
\dfrac{\pi}{2} \geq c_1 \geq c_2 \geq c_3 \geq 0
\end{equation*}
and
\begin{equation}\label{eq11}
\pi \geq c_1 > \dfrac{\pi}{2},~\pi - c_1 \geq c_2 \geq c_3 \geq 0.
\end{equation}

Each local equivalence class is represented by a unique point in the tetrahedron, with the exception that the points $(c_1, c_2, 0)$ and $(\pi - c_1, c_2, 0)$ represent the same local equivalence class. 

\begin{widetext}
\begin{center}
\begin{figure}[h]
\includegraphics[width=1.0\textwidth]{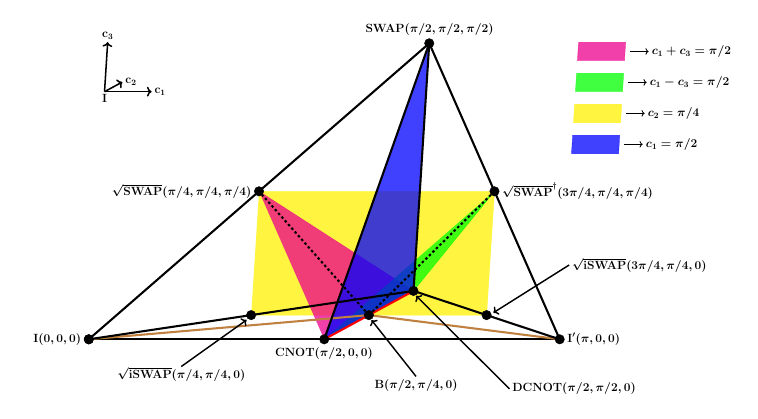}
\caption{Weyl chamber in the positive Cartan coordinate system. The planar regions describing the mirror operation are highlighted.}
\label{fig1}
\end{figure}
\end{center}
\end{widetext}

\subsection{Inverse and Mirror Operations in the Weyl Chamber}

In the Weyl chamber, the change of Cartan coordinates of two-qubit gates upon multiplication with the SWAP gate and the Hermitian conjugate operation can be described as reflections with respect to certain planes passing through the Weyl chamber. If the two-qubit gate $U$ belongs to the local equivalence class represented by the point $(c_1, c_2, c_3)$ in the Weyl chamber, then its inverse $U^\dagger$ belongs to the local equivalence class represented by the point $(\pi - c_1, c_2, c_3)$. Thus, the inverse operation corresponds to the reflection of Cartan coordinates with respect to the $c_1 = \pi/2$ plane of the Weyl chamber, and the one-half of the Weyl chamber is the inverse of its other half. The mirror gate of $U$, which is $U.\text{SWAP}$ or $\text{SWAP}.U$ belongs to the local equivalence class represented by the point $\left( \dfrac{\pi}{2} + \text{sgn} \left( \dfrac{\pi}{2} - c_1 \right) c_3, \dfrac{\pi}{2} - c_2, \text{sgn} \left( \dfrac{\pi}{2} - c_1 \right) \left( \dfrac{\pi}{2} - c_1 \right) \right)$~\cite{Peterson2022,Cross2019}, where the function $\text{sgn}(x) = +1$, if $x \geq 0$; else $\text{sgn}(x) = -1$. This change of Cartan coordinates can be described as reflections about $c_1 = \pi/2$, $c_2 = \pi/4$, and $((\pi/2) - c_1) - \text{sgn}((\pi/2) - c_1)c_3 = 0$ planes of the Weyl chamber. The last plane is either $c_1 + c_3 = \pi/2$ or $c_1 - c_3 = \pi/2$ depending on the value of $\text{sgn}((\pi/2) - c_1)$. The reflecting planes are highlighted in FIG.~\ref{fig1}. 

The corresponding reflecting planes in the balanced Cartan coordinate system were described, and the set of reflections was defined as the mirror operation in Ref.~\cite{Selvan2023}. Both a two-qubit gate and its mirror gate have the same ability to generate other two-qubit gates. If the two-qubit gate $U$ generates another two-qubit gate $V$ in two applications, then the mirror gate of $U$ can also generate $V$ in two applications~\cite{Zhang2005}.

\subsection{Nonlocal Content of Product of Two-Qubit Gates}

\textit{Nonlocal content:} For the two-qubit gate $U$, given in Eq.~\ref{eq1}, we define the nonlocal content of $U$, $\phi(U)$, as the vector $\lambda(H)$ given below. 
\begin{equation}\label{eq12}
\phi(U) =  \lambda (H) = \left(a_1 = \dfrac{h_2}{2 \pi}, a_2 = \dfrac{h_1}{2 \pi}, a_3 = \dfrac{h_4}{2 \pi}, a_4 = \dfrac{h_3}{2 \pi} \right),
\end{equation}
where $h_j$'s $(j=1,2,3,4)$ are the eigenvalues of $H$ given in Eqs.~\ref{eq4} - \ref{eq7}. 

The elements of $\phi(U)$ satisfy the following conditions.
\begin{equation*}
a_1 \geq a_2 \geq a_3 \geq a_4,~~a_1 - a_4 \leq 1,
\end{equation*}
and 
\begin{equation}\label{eq13}
a_1 + a_2 + a_3 + a_4 = 0.
\end{equation}

In Refs.~\cite{Zhang2005}, the eigenvalues of $H$ arranged in descending order were defined as the nonlocal content of $U$. But the definition given in Eq.~\ref{eq12} is convenient. 

The ability of a local equivalence class to generate two-qubit gates can be studied from the nonlocal content of the product of two-qubit gates from the local equivalence class. If $U_1$, $U_2$ and $U_3$ are two-qubit gates with nonlocal contents $\phi(U_1) = [b_1, b_2, b_3, b_4]$, $\phi(U_2) = [e_1, e_2, e_3, e_4]$, and $\phi(U_3) = [f_1, f_2, f_3, f_4]$ such that $U_1 U_2 = U_3$, then for two integers $r, k > 0$ such that $r+k=4$, $\alpha, \beta, \delta \in \mathcal{Q}_{rk} = \{(I_1,...,I_r) | k \geq I_1 \geq,...,\geq I_r \geq 0  \}$, and $d \geq 0$ such that the quantum Littlewood–Richardson coefficient, $N_{\alpha \beta}^{\delta, d} = 1$, the following inequality relation holds~\cite{Peterson2020}.

\begin{equation}\label{eq14}
d - \sum_{j=1}^r b_{k+j-\alpha_j} - \sum_{j=1}^r e_{k+j-\beta_j} + \sum_{j=1}^r f_{k+j-\delta_j} \geq 0
\end{equation}

There are 74 sets of values of $\{r, k, \alpha, \beta, \delta, d\}$ for which $N_{\alpha \beta}^{\delta, d} = 1$~\cite{Peterson2020}. To find the two-qubit gates that can be generated, in the Weyl chamber [FIG.~\ref{fig1}], by the product of gates from the local equivalence classes of $U_1$ and $U_2$, the 74 inequality relations corresponding to the nonlocal contents of the four sets $\{ \pm U_1, \pm U_2\}$, with the conditions of Weyl chamber [Eq.~\ref{eq11}] imposed, have to be analyzed. It has to be noted that though $U$ and $-U$ belongs to the same local equivalence class, the nonlocal content of $-U$ can be written differently as $[a_3 + 0.5, a_4 + 0.5, a_1 - 0.5, a_2 -0.5]$, which corresponds to the Cartan coordinates $(\pi - c_1, c_2, -c_3)$, and still satisfies the conditions given in Eq.~\ref{eq13}. 

\section{Optimal Construction of Two-Qubit Gates}

\begin{figure}[h]
\includegraphics[scale=0.7]{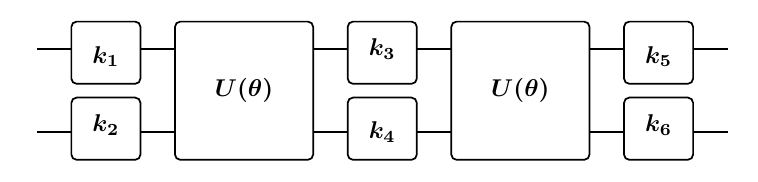}
\caption{Circuit for optimal generation of all two-qubit gates.}
\label{fig2}
\end{figure}

Our objective is to find the family of local equivalence classes to construct the circuit, shown in FIG.~\ref{fig2}, that can generate all two-qubit gates by changing the single-qubit gates, $k_i$'s $(i=1-6)$, and the two-qubit gate parameter $\theta$. By changing the parameter $\theta$, two-qubit gates from the local equivalence classes of the family can be generated. The infidelity of the circuit, due to dissipation, decoherence, and other noises, is a function of $\theta$. 

In general, a two-qubit gate $U$ from a local equivalence class is said to generate another two-qubit gate $V$ in two applications, if there exist two-qubit local gates $L_1$, $L_2$, and $L_3$ such that 
\begin{equation}
L_1 U L_2 U L_3 = V.
\label{eq15}
\end{equation}

This equation can be rewritten as either
\begin{equation}
L_1 U L_2 = V L_3^\dagger U^\dagger
\label{eq16}
\end{equation}
or 
\begin{equation}
L_2U L_3 = U^\dagger L_1^\dagger V.
\label{eq17}
\end{equation}

If $V$ is taken to be a two-qubit local gate, then the Eqs.~\ref{eq16} and \ref{eq17} suggest that for $U$ to generate a local gate in two applications, both $U$ and $U^\dagger$, should belong to the same local equivalence class~\cite{Zhang2005}. This implies the invariance of the local equivalence class under inverse operation in the Weyl chamber: 

\begin{equation}
(c_1, c_2, c_3) \rightarrow (\pi - c_1, c_2, c_3).
\label{eq18}
\end{equation}

Clearly, this condition is satisfied by $c_1 = \pi/2$ and $c_3 = 0$ planes of the Weyl chamber. If $V$ is taken to be the SWAP gate, then $U$ to generate the SWAP gate in two applications, both $U$ and the mirror of $U^\dagger$ should belong to the same local equivalence class. That is, to generate the SWAP gate in two applications, the local equivalence class should be invariant under the combined inverse and mirror operations: 

\begin{equation*}
(c_1, c_2, c_3) \rightarrow 
\end{equation*}
\begin{equation}
\left( \dfrac{\pi}{2} -\text{sgn} \left( \dfrac{\pi}{2} -c_1 \right)c_3, \dfrac{\pi}{2} - c_2, \text{sgn}\left(\dfrac{\pi}{2} - c_1 \right) \left(\dfrac{\pi}{2} - c_1 \right) \right).
\label{eq19}
\end{equation}

Local equivalence classes that are invariant under this transformation are represented by the two lines connecting the B gate equivalence class with the $\sqrt{\text{SWAP}}$ and $\sqrt{\text{SWAP}}^\dagger$ equivalence classes~\cite{Lin2022} (the dashed lines in FIG.\ref{fig1}). The local equivalence class that can generate both the two-qubit local gates and the SWAP should be invariant under the mirror operation. The B gate equivalence class is the only equivalence class that is invariant under mirror operation, and it can generate all two-qubit gates in two applications~\cite{Zhang2004}.

\subsection{One-Parameter Family with B Gate Equivalence Class}

It is obvious that any one-parameter family of local equivalence classes containing the B gate local equivalence class as a member can be used to construct the circuit, shown in FIG.~\ref{fig2}, to generate all two-qubit gates. Such parameterized circuits were constructed using the special perfect entanglers (SPEs) from the CNOT gate to the B gate, and from the B gate to the DCNOT gate~\cite{Selvan2023}. We show the ability of local equivalence classes of two specific one-parameter families to generate the two-qubit gates. The first one is with the Cartan coordinates $(c_1, \pi/4, \pi/2-c_1)$, where $c_1 \in [\pi/4, \pi/2]$, and the other is with the Cartan coordinates $(c_1, c_1/2, 0)$, where $c_1 \in [0, \pi/2]$. In FIG.~\ref{fig1}, the former family is represented by the dashed line connecting the $\sqrt{\text{SWAP}}$ and B gate equivalence classes and the latter is represented by the line in brown colour connecting the points I(0, 0, 0) and B$(\pi/2, \pi/4, 0)$, which is locally equivalent to the brown coloured line connecting the points B$(\pi/2, \pi/4, 0)$ and $\text{I}'(\pi, 0, 0)$.   

All the local equivalence classes in the family with Cartan coordinates $(c_1, \pi/4, (\pi/2) - c_1)$ can generate the SWAP gate in two applications, but only the B gate equivalence class can generate the two-qubit local gates in two applications. The region of the Weyl chamber that can be covered by the two applications of gates from each local equivalence class of the family is found using the inequality relations given in Eq.~\ref{eq14} and the lrs software~\cite{Avis2000}. The fractional volume of the Weyl chamber covered in two applications is plotted for eleven different values of $c_1$, in FIG.~\ref{fig3}a. The fractional volume of the Weyl chamber covered increases with $c_1$. The local equivalence class corresponding to $c_1 = \pi/4$ is the $\sqrt{\text{SWAP}}$ equivalence class. It can generate only the gates belonging to the local equivalence classes represented by the line connecting the SWAP and the CNOT equivalence classes. Hence, the fractional volume is zero. The B gate equivalence class corresponds to $c_1 = \pi/2$, and it generates all two-qubit gates. The regions of the Weyl chamber covered in two applications are shown for $c_1 = 2\pi/7, \pi/3,$ and $ 2 \pi/5$ in FIG.~\ref{fig4}a, \ref{fig4}b, and \ref{fig4}c, respectively. These results are also valid for the family of local equivalence classes with Cartan coordinates $(\pi - c_1, \pi/4, (\pi/2) - c_1)$, where $c_1 \in [\pi/4, \pi/2]$, represented by the line connecting the B gate equivalence class with the $\sqrt{\text{SWAP}}^\dagger$ equivalence class.  

In the Weyl chamber, the $c_3 = 0$ plane is the mirror of the $c_1 = \pi/2$ plane, which describes the inverse operation and also belongs to the planes describing the mirror operation in the positive Cartan coordinate system. Hence, the ability of a local equivalence class, in $c_1=\pi/2$ plane, to generate two-qubit gates in two applications can be learned by studying the ability of its mirror in $c_3 = 0$ plane. We study the ability of the local equivalence classes in the family with Cartan coordinates $(c_1, c_1/2, 0)$ for $c_1 \in [0, \pi/2]$ to generate two-qubit gates. We refer to this family as B$^\alpha$ family, where $\alpha \in [0, 1]$. They are the mirror of the local equivalence classes represented by the line connecting the SWAP and B gate equivalence classes. The fractional volume of the Weyl chamber covered in two applications increases with $c_1$ as shown in FIG.~\ref{fig3}b. The region of Weyl chamber covered in two applications for $c_1 = \pi/8$, $\pi/4$, and $3\pi/4$ are shown in FIGs.~\ref{fig4}d, \ref{fig4}e, and \ref{fig4}f, respectively. All the local equivalence classes in this family can generate the two-qubit local gates in two applications. But none of them, except the B gate equivalence class, can generate the SWAP gate in two applications. 

\begin{widetext}
\begin{center}
\begin{figure}[h]
\begin{tabular}{c c}
\includegraphics[width=0.5\textwidth]{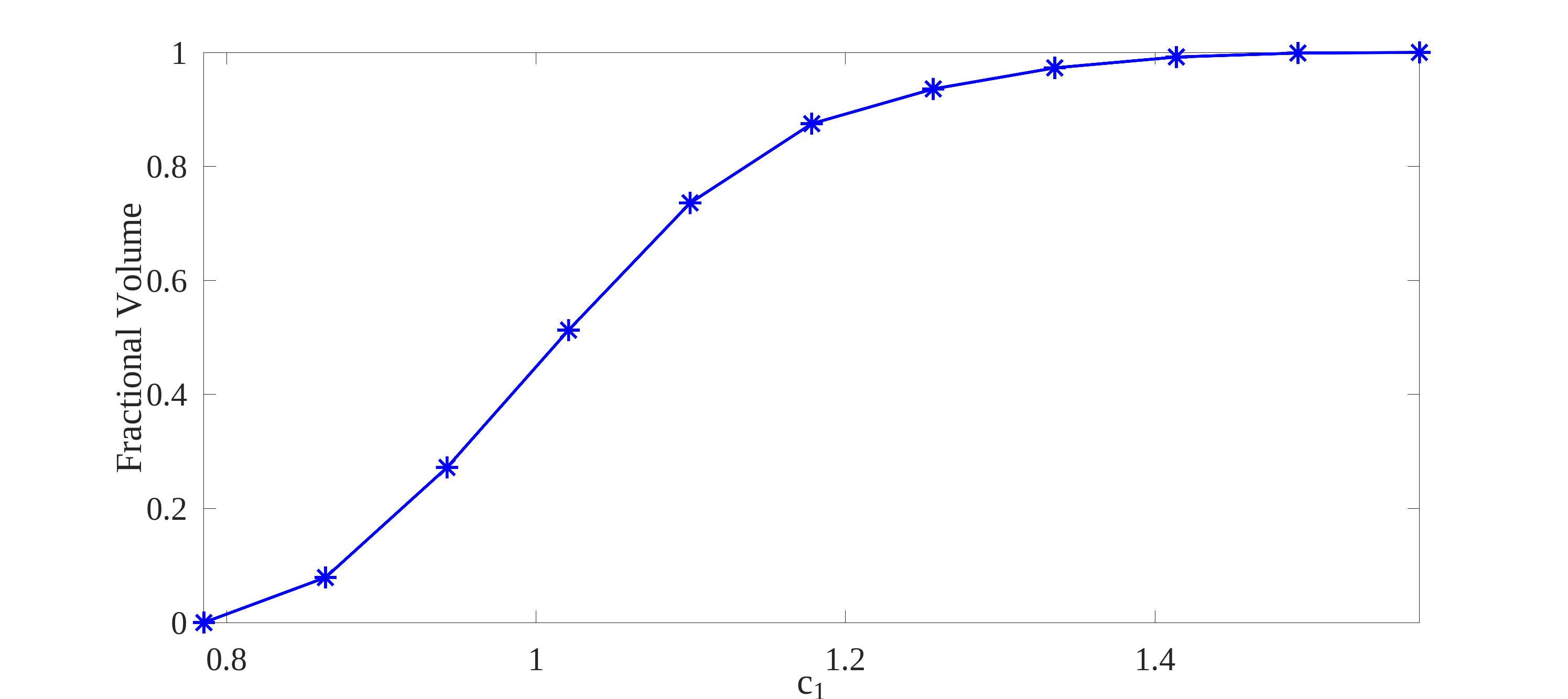} & \includegraphics[width=0.5\textwidth]{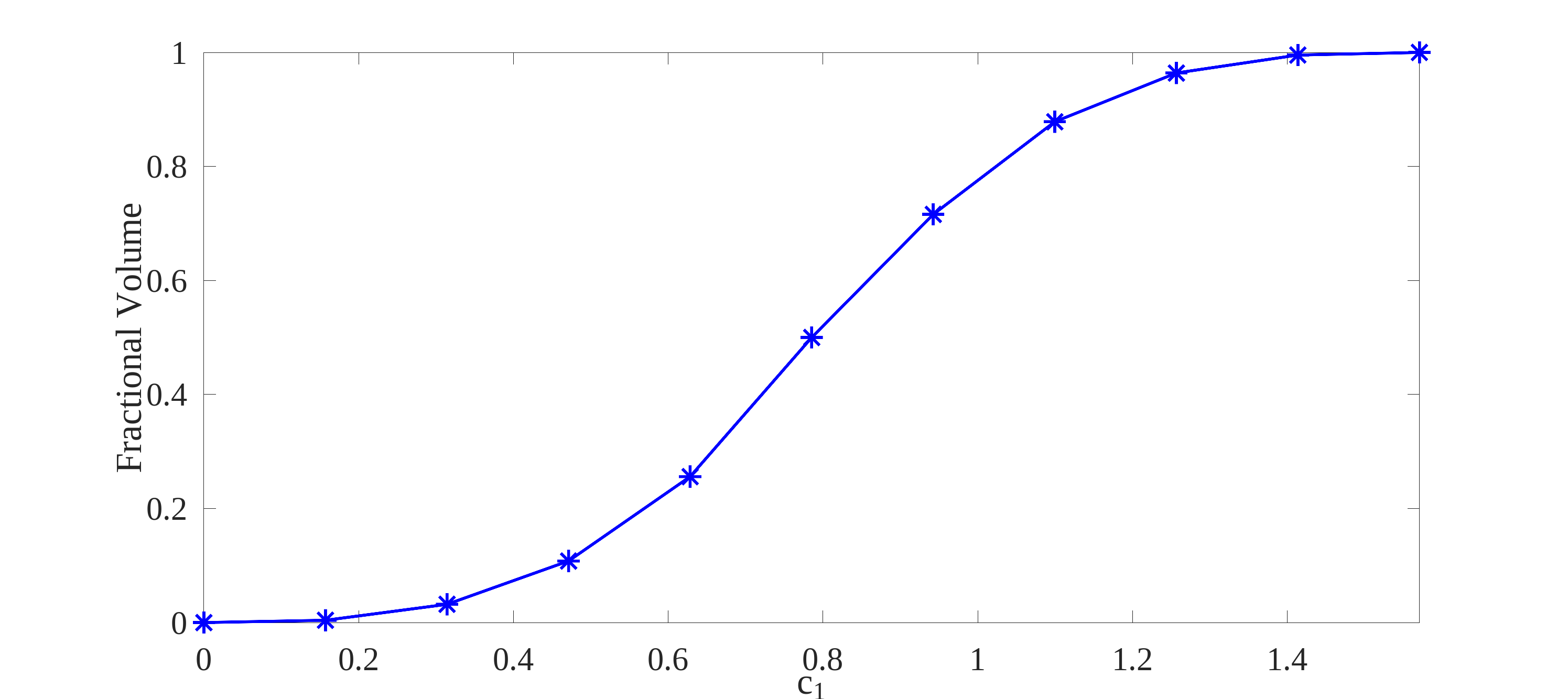}  \\ (a) & (b)  
\end{tabular}
\caption{Fractional volume of Weyl chamber covered vs $c_1$ for the families with Cartan coordinates (a) $(c_1, \pi/4, (\pi/2) - c_1)$ for eleven equally spaced values of $c_1$ from $\pi/4$ to $\pi/2$, and (b) $(c_1, c_1/2, 0)$ for eleven equally spaced values of $c_1$ from $0$ to $\pi/2$.}
\label{fig3}
\end{figure}
\end{center}
\end{widetext}

In the two families discussed above, the region of the Weyl chamber covered by two applications of gates from a local equivalence class with $c_1 = c$, for some value of $c$, is also covered by all the local equivalence classes with $c_1 > c$. Hence, these families are useful for quantum computation only if a gate from the local equivalence class with $c_1 = c$ can be implemented with better fidelity than the gates belonging to the local equivalence classes with $c_1 > c$. From this perspective, the B$^\alpha$ family is suitable for quantum computation. The gates from the local equivalence class I$(0, 0, 0)$ to B$(\pi/2, \pi/4, 0)$ can be generated by increasing the time of evolution of qubits under the Hamiltonian $2g(\sigma_x \otimes \sigma_x)+g(\sigma_y \otimes \sigma_y)$ with constant interaction strength, $g$. In general, the infidelity of the gates increases with implementation time. 

In $c_3 = 0$ and $c_1 = \pi/2$ planes, it is not possible to have a family of local equivalence classes without the B gate equivalence class that can be used to construct the circuit, shown in FIG.~\ref{fig2}, to optimally generate all two-qubit gates. However, it is possible to have families of local equivalence classes, without the B gate equivalence class, in the $c_1 \pm c_3 = \pi/2$ and $c_2 = \pi/4$ planes. Because these two planes contain the local equivalence classes that generate the SWAP gate and the two-qubit local gates in two applications.

\begin{widetext}
\begin{center}
\begin{figure}[h]
\begin{tabular}{c c c}
\includegraphics[width=0.3\textwidth]{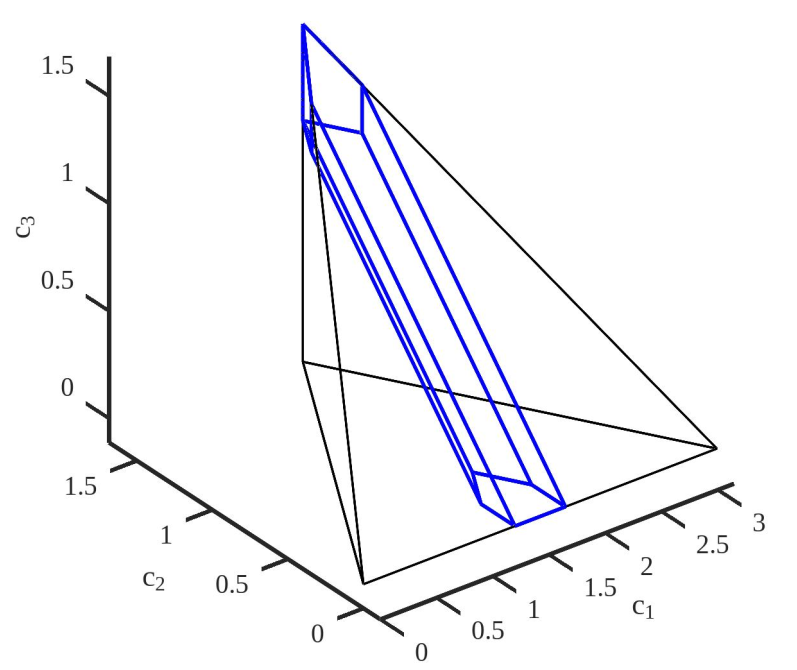} & \includegraphics[width=0.3\textwidth]{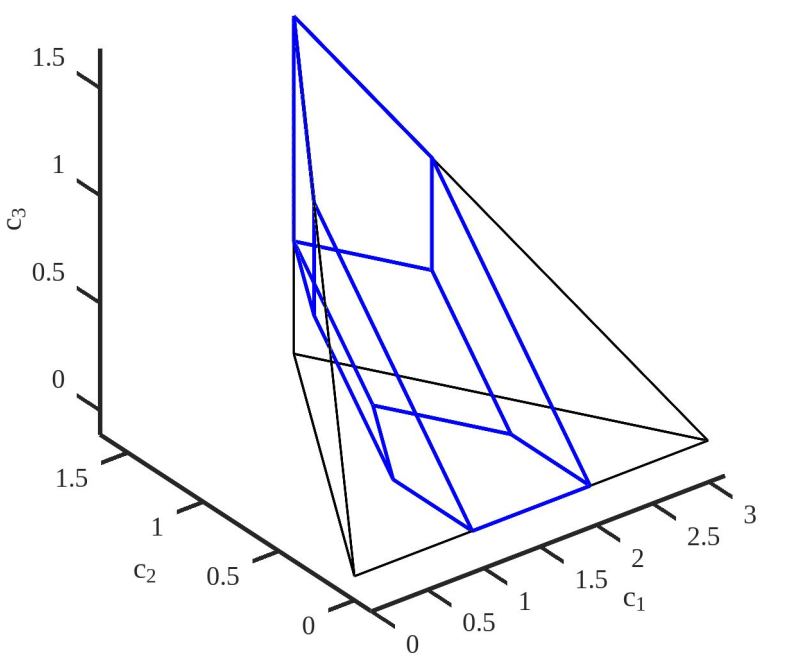} & \includegraphics[width=0.3\textwidth]{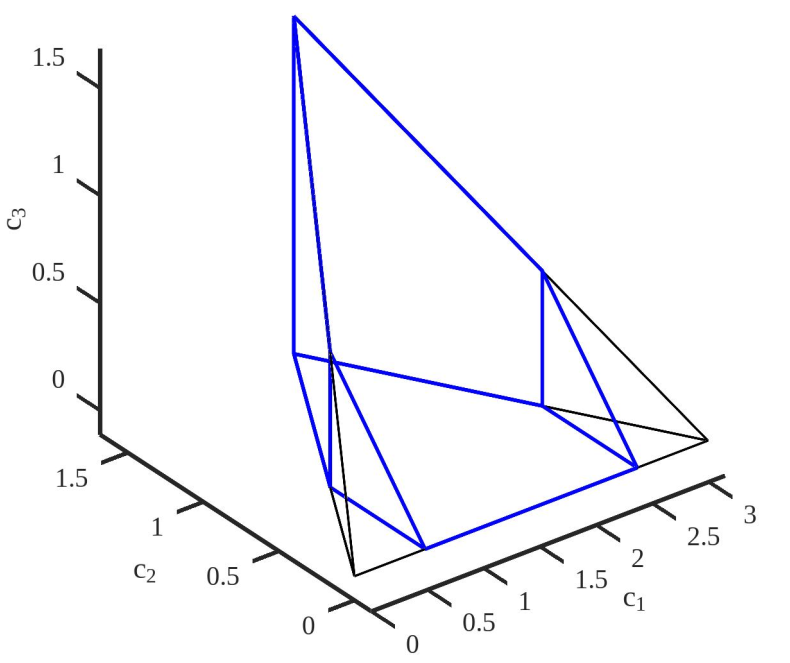} \\ (a) & (b) & (c) \\ & & \\ \includegraphics[width=0.3\textwidth]{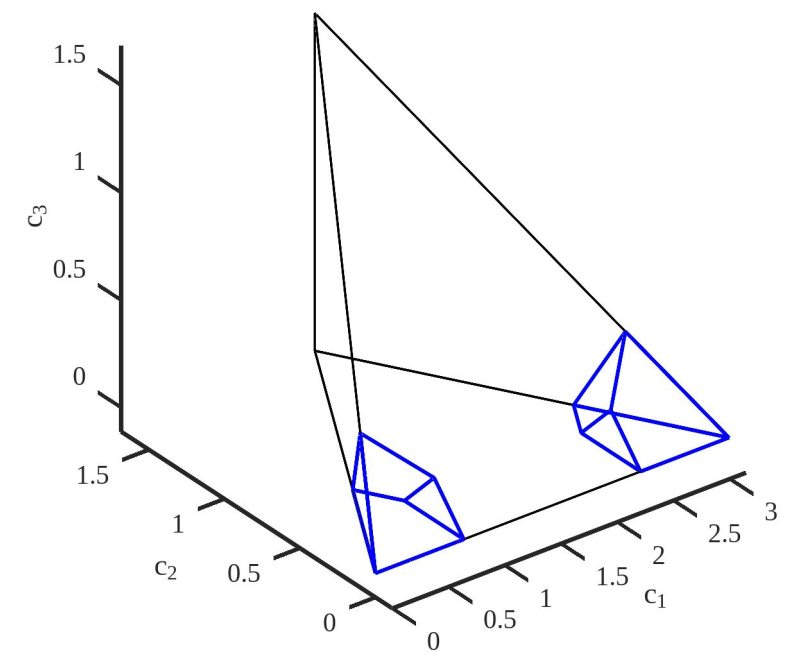} & \includegraphics[width=0.3\textwidth]{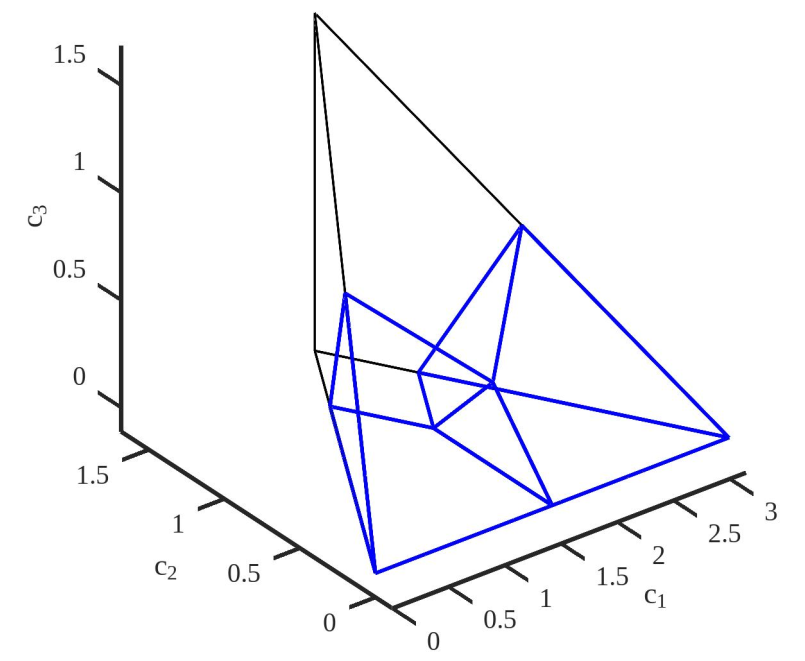} & \includegraphics[width=0.3\textwidth]{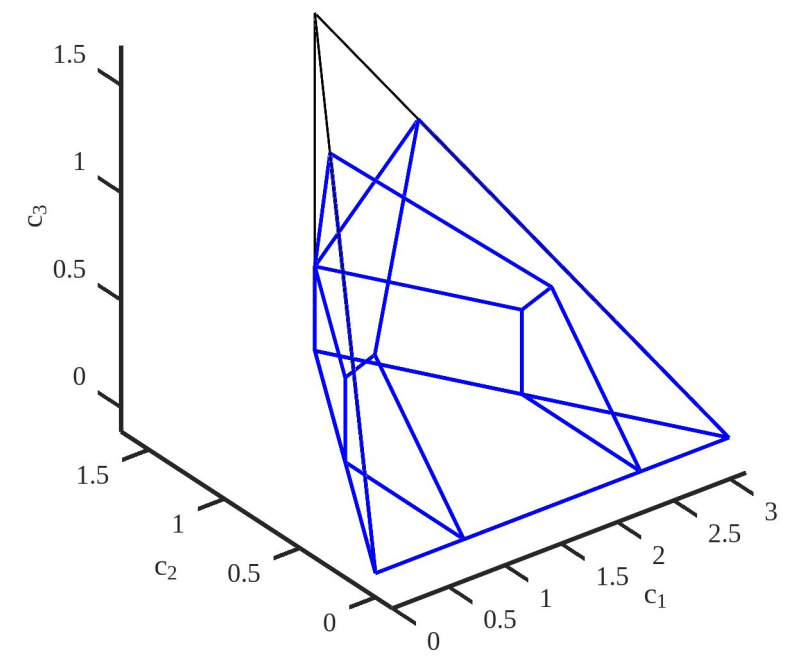} \\ (d) & (e) & (f)
\end{tabular}
\caption{The region of Weyl chamber covered by two applications of gates from the local equivalence class with Cartan coordinates (a) $(2\pi/7, \pi/4, 3\pi/14)$, (b) $(\pi/3, \pi/4, \pi/6)$, (c) $(2\pi/5, \pi/4, \pi/10)$, (d) $(\pi/8, \pi/16, 0)$, (e) $(\pi/4, \pi/8, 0)$, and (f) $(3\pi/4, 3\pi/8, 0)$. }
\label{fig4}
\end{figure}
\end{center}
\end{widetext}

\subsection{One-Parameter Family without B Gate Equivalence Class}

First, we consider $c_1 + c_3 = \pi/2$ triangular region of Weyl chamber, shown in FIG.~\ref{fig5}a. It contains the line representing the SPEs and the line connecting $\sqrt{\text{SWAP}}$ and the B gate equivalence class, which bisects the triangular region into two right-angled triangles. The mirror of the inverse (or the inverse of the mirror) of the local equivalence class represented by the point $(c_1, c_2, (\pi/2) - c_1)$ on the left side right-angled triangle is represented by the point $(c_1, (\pi/2) - c_2, (\pi/2) - c_1)$ on the right side right-angled triangle. Each SPE equivalence class can generate the two-qubit local gates in two applications, and the local equivalence class bisecting the $c_1 + c_3 = \pi/2$ triangular region can generate the SWAP gate in two applications. Hence, we consider the family of local equivalence classes represented by lines parallel to the hypotenuse and connecting points on the opposite and adjacent sides of the right-angled triangle. On the left side right-angled triangle, each line segment can be labelled by an angle $\theta \in [0, \pi/4]$ such that the Cartan coordinates of the local equivalence classes of a family represented by the line segment are $((\pi/2) + \theta - c_2, c_2, c_2 - \theta)$, where $c_2 \in [\theta, \pi/4]$. The Cartan coordinates of the corresponding mirrored inverse local equivalence classes are $((\pi/2) + \theta - c_2, (\pi/2) - c_2, c_2 - \theta)$.

Two such lines for $\theta = \pi/6~\text{and}~\pi/12$ are shown in FIG.~\ref{fig5}a. To show that the family of local equivalence classes represented by each of these lines can be used to form the circuit, shown in FIG.~\ref{fig2}, to generate all two-qubit gates, we find the regions of the Weyl chamber covered by two applications of the gates from the local equivalence classes represented by the points marked on the lines. The regions of the Weyl chamber covered by the marked points are shown in FIGs.~\ref{fig6}a and~\ref{fig6}b for $\theta = \pi/6~\text{and}~\pi/12$ respectively. It is evident that their union covers the entire Weyl chamber. 

The region of the Weyl chamber covered in two applications for $c_2 = \pi/16, \pi/8, 3\pi/16$ on the line with $\theta = 0$ and the corresponding mirrored inverse points on the other side of the triangular region are shown in FIG.~\ref{fig6}c. In FIG.~\ref{fig6}c, for smaller $c_2$, the covered region of the Weyl chamber is wider and shorter. For $c_2 = 0$, we have the CNOT equivalence class on one side and the DCNOT equivalence class on the other side. Both of them can only generate all the two-qubit gates represented by the $c_3 = 0$ plane of the Weyl chamber in two applications~\cite{McKinney2023}. The $\sqrt{\text{SWAP}}$ equivalence class with $c_2 = \pi/4$ can only generate the gates represented by the line connecting the points representing the SWAP and the CNOT equivalence classes, in two applications. We can expect continuous change from the $c_3 = 0$ plane to the $c_1 = \pi/2$ line on the $c_2 = c_3$ plane of the Weyl chamber as $c_2$ is continuously varied from 0 to $\pi/4$. Hence, we conjecture that a family of local equivalence classes with Cartan coordinates $((\pi/2) + \theta - c_2, c_2, c_2 - \theta)$, for given $\theta$, either alone or along with the family consisting of their mirrored inverse local equivalence classes can generate all the two-qubit gates in two applications as described by the circuit shown in FIG.~\ref{fig2}. Since both $c_1 \pm c_3 = \pi/2$ planes are inverse of each other, a similar conjecture can also be made for the $c_1 - c_3 = \pi/2$ plane. 

\begin{widetext}
\begin{center}
\begin{figure}[h]
\begin{tabular}{c c}
\includegraphics[width=0.5\textwidth]{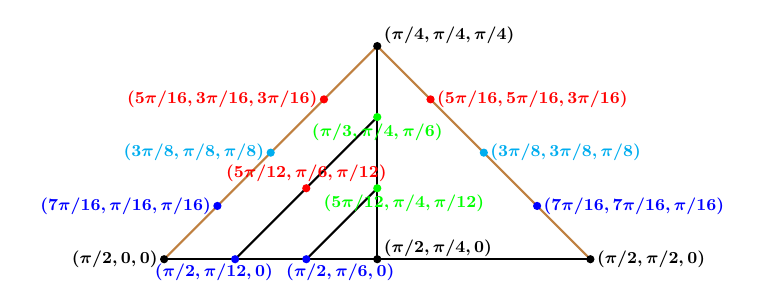} & \includegraphics[width=0.5\textwidth]{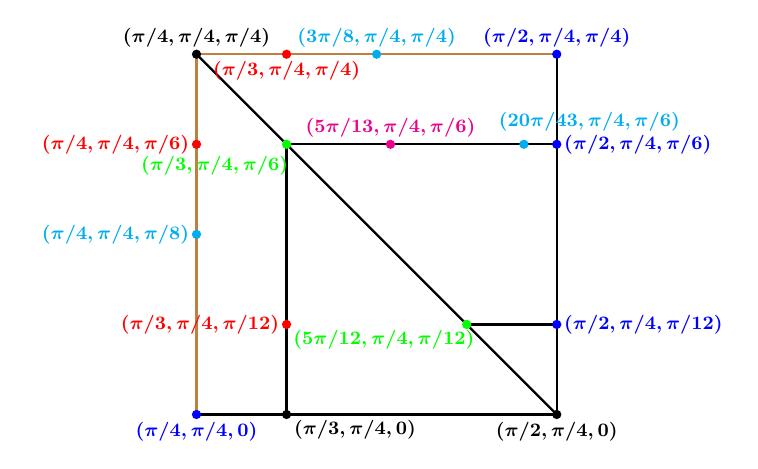}  \\ (a) & (b) 
\end{tabular}
\caption{(a) $c_1 + c_3 = \pi/2$ planar region of Weyl chamber, and (b) the first half of $c_2 = \pi/4$ planar region of Weyl chamber. In both subfigures, a point and its Cartan coordinates are in the same color. }
\label{fig5}
\end{figure}
\end{center}
\end{widetext}

Now, we consider the first half of $c_2 = \pi/4$ plane of the Weyl chamber, shown in FIG.~\ref{fig5}b. The local equivalence classes represented by the diagonal can generate the SWAP gate in two applications, and the local equivalence classes represented by $c_1 = \pi/2$ and $c_3 = 0$ line segments can generate the two-qubit local gates in two applications. A horizontal line connecting a point on the $c_1 = \pi/2$ with a point on the diagonal, and the vertical line connecting the same point on the diagonal with a point on the $c_3 = 0$ line are mirrored inverse of each other. Two horizontal lines corresponding to $c_3 = \pi/12~\text{and}~\pi/6$ are shown in FIG.~\ref{fig5}b. The regions of the Weyl chamber covered by two applications of the gates from the local equivalence classes represented by the points marked on the $c_3 = \pi/12$ and $c_3 = \pi/6$ (and its mirrored inverse line) lines are shown in FIGs.~\ref{fig6}d and~\ref{fig6}e, respectively. The same is shown in FIG.~\ref{fig6}f for the local equivalence classes represented by the points marked on $c_3 = \pi/4$ line and their mirrored inverse points on $c_1 = \pi/4$ line. 

The entire Weyl chamber is covered in FIG.~\ref{fig6}d and almost covered in FIG.~\ref{fig6}e. In FIG.~\ref{fig6}f, the entire Weyl chamber is not covered. For the Cartan coordinates $(\pi/2, \pi/4, \pi/4)$ and $(\pi/4, \pi/4, 0)$, the region of the Weyl chamber covered is a tetrahedron (blue). For $(\pi/4, \pi/4, \pi/4)$, the region is a line segment from the SWAP to the CNOT equivalence classes, as mentioned earlier. From the shape of the other two regions, one can expect that the entire region of the Weyl chamber can be covered by the union of regions corresponding to all the local equivalence classes represented by $c_1 = \pi/4$ and $c_3 = \pi/4$ line segments, shown in FIG.~\ref{fig5}b. Hence, we conjecture that a family of local equivalence classes represented by a horizontal line connecting a point on $c_1 = \pi/2$ line with a point on the diagonal, in FIG.~\ref{fig5}b, either alone or together with its mirrored inverse, can generate all the two-qubit gates in two applications as described by the circuit shown in FIG.~\ref{fig2}. A similar statement can also be made for the other half of the $c_2 = \pi/4$ plane of the Weyl chamber.  

\begin{widetext}
\begin{center}
\begin{figure}[h]
\begin{tabular}{c c c}
\includegraphics[width=0.3\textwidth]{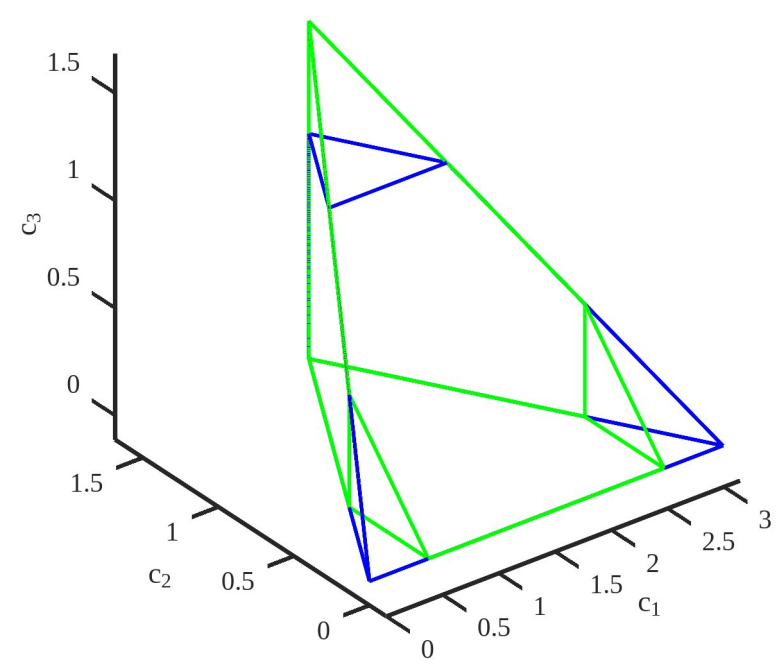} & \includegraphics[width=0.3\textwidth]{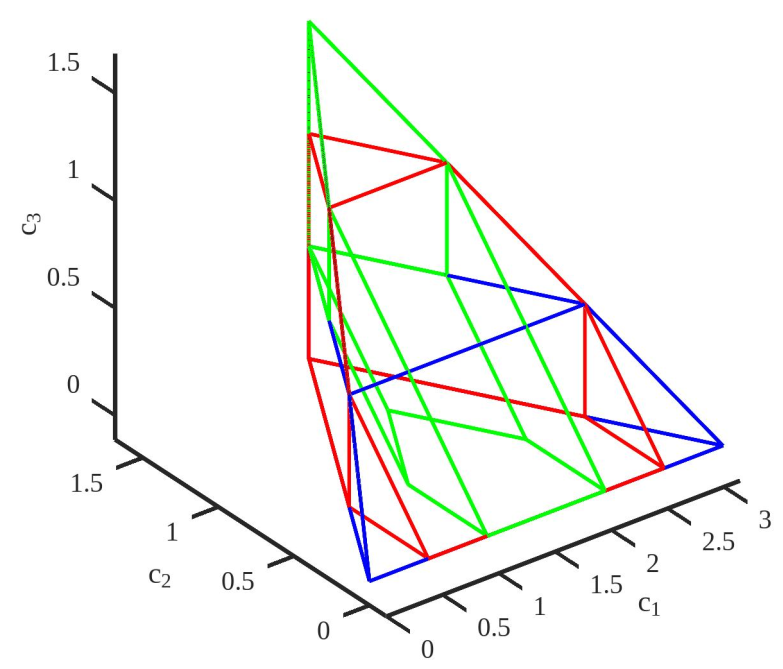} & \includegraphics[width=0.3\textwidth]{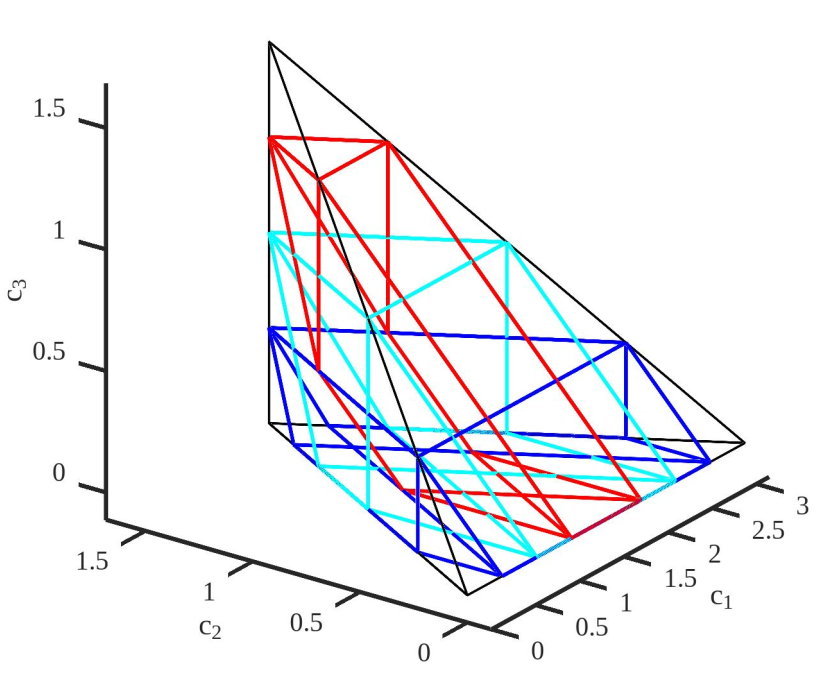}\\ (a) & (b) & (c) \\ & \\ \includegraphics[width=0.3\textwidth]{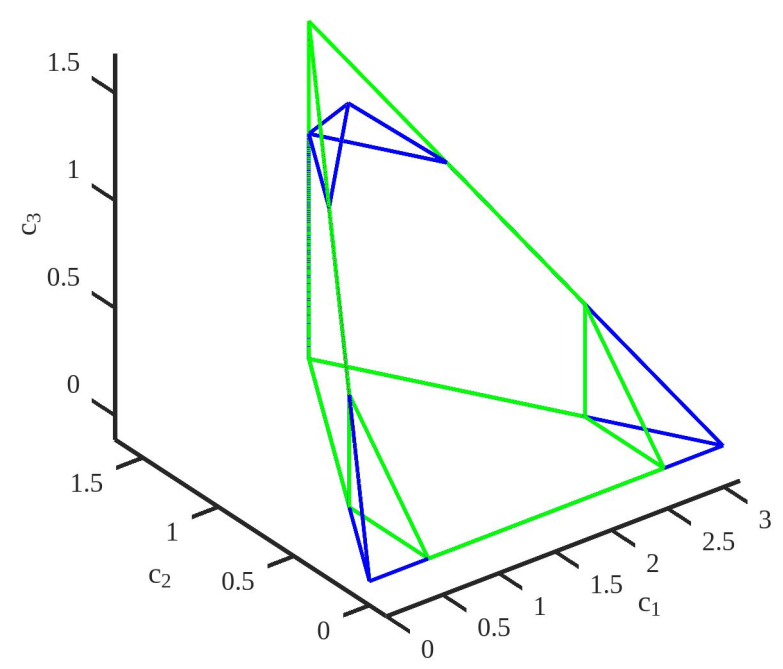} & \includegraphics[width=0.3\textwidth]{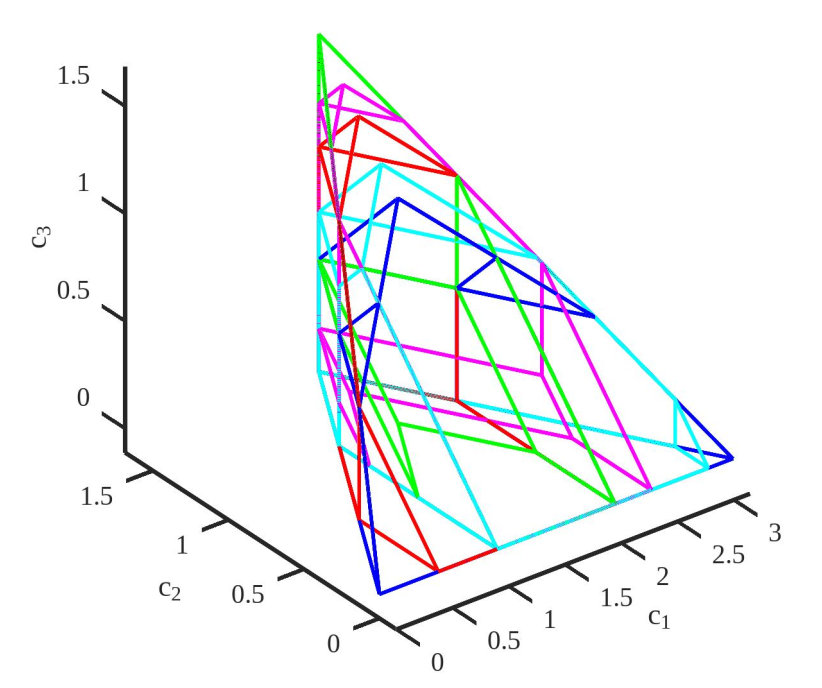} & \includegraphics[width=0.3\textwidth]{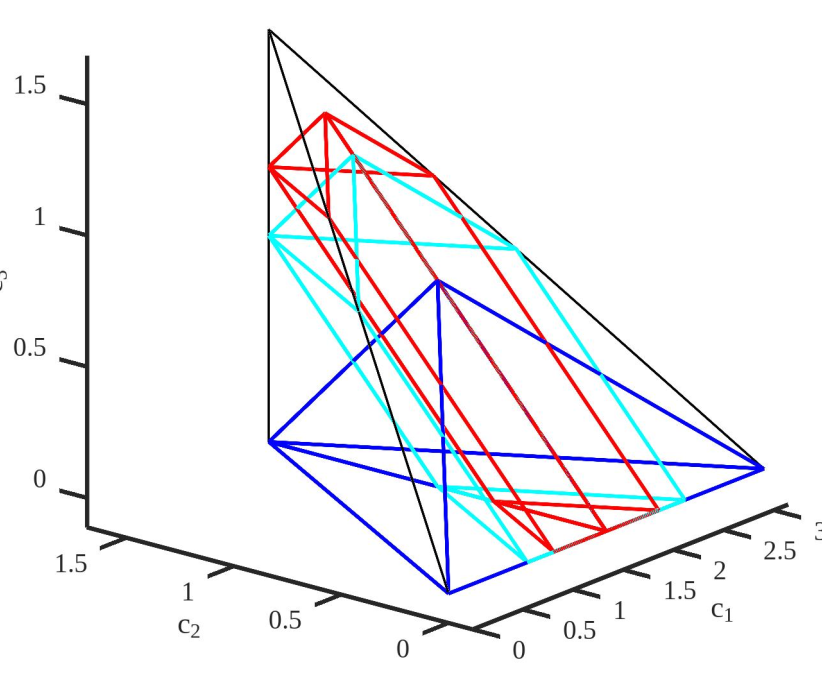}\\ (d) & (e) & (f)
\end{tabular}
\caption{The polytope region described by the set of 74 inequality relations [Eq.~\ref{eq14}] corresponding to the pair $\{+U, +U\}$, where $U$ is a gate from the local equivalence class with Cartan coordinates (a) $(\pi/2, \pi/6, 0)$ (blue colour) and $(5\pi/12, \pi/4, \pi/12)$ (green colour), (b) $(\pi/2, \pi/12, 0)$ (blue colour), $(5\pi/12, \pi/6, \pi/12)$ (red colour), and $(\pi/3, \pi/4, \pi/6)$ (green color), (c) $(7\pi/16, \pi/16, \pi/16)$ and its mirrored inverse (blue colour), $(3\pi/8, \pi/8, \pi/8)$ and its mirrored inverse (cyan colour), and $(5\pi/16, 3\pi/16, 3\pi/16)$ and its mirrored inverse (red colour), (d) $(\pi/2, \pi/4, \pi/12)$ (blue colour), and $(5\pi/12, \pi/4, \pi/12)$ (green colour), (e) $(\pi/2, \pi/4, \pi/6)$ (blue colour), $(20\pi/43, \pi/4, \pi/6)$ (cyan colour), $(\pi/3, \pi/4, \pi/12)$ (red color), $(5\pi/13, \pi/4, \pi/6)$ (magenta colour), and $(\pi/3, \pi/4, \pi/6)$ (green colour), and (f) $(\pi/2, \pi/4, \pi/4)$ and its mirrored inverse (blue colour), $(3\pi/8, \pi/4, \pi/4)$ and its mirrored inverse (cyan colour), and $(\pi/3, \pi/4, \pi/4)$ and its mirrored inverse (red colour). The region described by the set of 74 inequality relations [Eq.~\ref{eq14}] corresponding to the pair $\{+U, -U\}$  is shown only in the subfigures (d) for $(\pi/2, \pi/4, \pi/12)$ (blue colour) and (e) for $(\pi/2, \pi/4, \pi/6)$ (blue colour).}
\label{fig6}
\end{figure}
\end{center}
\end{widetext}

\section{Implementation of Optimal Circuit}

To use the optimal circuit, shown in FIG.~\ref{fig2}, for quantum computation, schemes are required to implement the gates belonging to the family of local equivalence classes, discussed in Section III. All of them involve either two or all three of the $XX$, $YY$, and $ZZ$ interaction terms. Among them, the entire B$^\alpha$ family requires a relatively simple interaction $(2XX+YY)$ for implementation. This term can be realized using a combination of different interactions. Implementation of the gates belonging to the B gate equivalence class between two superconducting qubits has been reported~\cite{Chen2025,Wei2024,Selvan2024}. In Ref.~\cite{Chen2025}, a gate belonging to the B gate equivalence class was implemented between two superconducting qubits with a tunable coupler by driving one of the qubits in addition to the exchange interaction between the qubits. However, this scheme may require a longer implementation time for gates belonging to the B$^\alpha$ family with $\alpha$ close to 0. But, the gates that are implementable using the scheme presented in Ref~\cite{Chen2024}, from the rest of the equivalence classes, can form a continuous basis gate set for optimal generation of all two-qubit gates. 

\begin{figure}[h]
\includegraphics[scale=0.5]{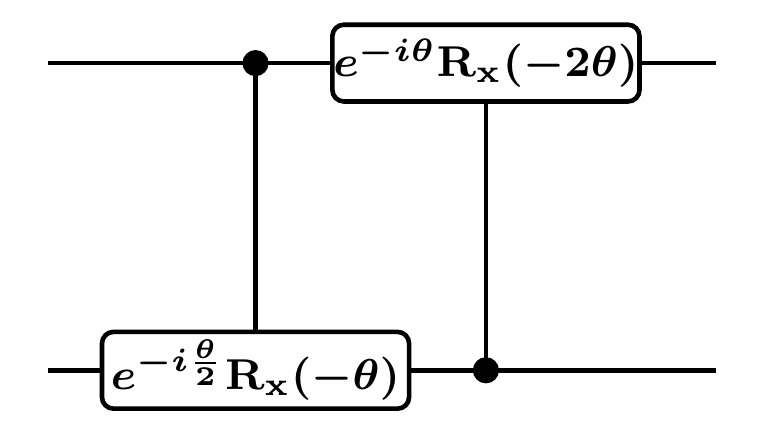}
\caption{Parameterized circuit generating the gates belonging to the local equivalence classes of the B$^\alpha$ family.}
\label{fig7}
\end{figure}

In Ref.~\cite{Selvan2024}, a circuit locally equivalent to the B gate was implemented between two superconducting qubits with fixed coupling using a sequence of cross-resonance pulses. A parameterized two-qubit circuit generating the gates belonging to the local equivalence classes of the B$^\alpha$ family is shown in FIG.~\ref{fig7}, where $R_x(-\theta) = \exp(i \theta \sigma_x/2)$. The circuit parameter $\theta$ takes values from $0$ to $\pi/2$. This circuit can be implemented in superconducting quantum processors, where the cross-resonance interaction is the native interaction, using the pulse-level control techniques~\cite{Earnest2021,Stenger2021}. However, the optimal circuit, shown in FIG.~\ref{fig2}, with $U(\theta)$ replaced by the circuit shown in FIG.~\ref{fig7}, will take more time to implement many two-qubit gates than their implementation using the echoed cross-resonance (ECR) gate.

In Ref.~\cite{Wei2024}, the nonlocal part of the B gate was implemented between two superconducting qubits with fixed coupling by combining the iSWAP$^{3/4}$ and bSWAP$^{1/4}$ gates. This scheme can be extended to implement the nonlocal parts of the entire B$^\alpha$ family. It requires the combination of iSWAP$^{3\alpha/4}$ and bSWAP$^{\alpha/4}$ gates. In this scheme, the implementation time can be expected to increase with $\alpha$, and hence this scheme can be used for optimal generation of two-qubit gates using the B$^\alpha$ family. 

Two-qubit gates belonging to other families of local equivalence classes discussed in Section III require all three $XX$, $YY$, and $ZZ$ interaction terms (except the SPEs). Those gates can be implemented by combining the cross-resonance interaction with XX+YY and XX-YY interactions, as mentioned in Ref.~\cite{Wei2024}. The families of local equivalence classes on $c_2 = \pi/4$ and $c_1 + c_3 = \pi/2$ planes, represented by the brown coloured lines in FIG.~\ref{fig5}, contain the gates belonging to the fermionic simulation (fSim) gate set~\cite{Foxen2020}. This gate set involves two parameters $\theta$ and $\phi$. It is represented by the following matrix. 

\begin{equation}
\text{fSim}(\theta, \phi) = \begin{bmatrix}
1 & 0 & 0 & 0 \\ 0 & \cos (\theta) & -i \sin (\theta) & 0 \\ 0 & -i \sin (\theta) & \cos (\theta) & 0 \\ 0 & 0 & 0 & e^{- i \phi} 
\end{bmatrix}
\end{equation}

The local invariants of this matrix are

\begin{equation}
G_1 = \dfrac{1}{4} \left[ 2 \cos( 2 \theta) + \dfrac{[3 + \cos (4 \theta) ] \cos(\phi)}{2} + i \sin^2(2 \theta) sin(\phi) \right], 
\end{equation}
and
\begin{equation}
G_2 = 2 \cos( 2 \theta) + \cos(\phi). 
\end{equation}

These expressions correspond to the gates represented by the points on $c_1 = c_2$ and $c_2 = c_3$ planes of the Weyl chamber [FIG.~\ref{fig1}]. The gate set parameters $\theta = c_1 [c_3]$ and $\phi = 2c_3 [2c_1]$ represent a gate belonging to the local equivalence class with Cartan coordinates $(c_1, c_1, c_3)$ $[(c_1, c_3, c_3)]$. In FIG.~\ref{fig5}a, the brown coloured line connecting the points $(\pi/2, 0, 0)$ and $(\pi/4, \pi/4, \pi/4)$ [$(\pi/4, \pi/4, \pi/4)$ and $(\pi/2, \pi/2, 0)$] is on the $c_2 = c_3$ [$c_1 = c_2$] plane of the Weyl chamber. Similarly, in FIG.~\ref{fig5}b, the brown coloured line connecting the points $(\pi/4, \pi/4, 0)$ and $(\pi/4, \pi/4, \pi/4)$ [$(\pi/4, \pi/4, \pi/4)$ and $(\pi/2, \pi/4, \pi/4)$] is on the $c_1 = c_2$ [$c_2 = c_3$] plane of the Weyl chamber. These gates can be implemented with the suitable combinations of iSWAP-like (fSim$(\theta, \phi \propto \theta^2)$), and controlled phase (fSim$(\theta = 0, \phi)$) gates~\cite{Foxen2020}.  

\section{On Synthesis of n-Qubit Gates}

To use these families of fsim gates as a continuous basis gate set to perform quantum computation, their ability to generate an arbitrary $n$-qubit gate needs to be studied. It has to be noted that twice the lower bound $\left( \left\lceil \dfrac{4^n - 3n -1}{9} \right\rceil \right)$~\cite{Yu2013} on the number of two-qubit gates required to generate an arbitrary $n$-qubit gate is smaller than the lower bound $\left( \left\lceil \dfrac{4^n - 3n -1}{4} \right\rceil \right)$~\cite{Shende2006} on the number of CNOT gates required. Generally, for a given two-qubit gate, the upper bounds are obtained using the quantum Shannon decomposition~\cite{Shende2006}, by which an arbitrary $n$-qubit gate can be implemented by three multiplexed rotations and four general $(n-1)$-qubit gates. If $c_j$ is the least number of gates required to generate an arbitrary $j$-qubit gate, then for the CNOT gate, we have the following inequality relation~\cite{Shende2006}. 
\begin{equation}
 c_j \leq 4 c_{j-1} + 3 \times 2^{j-1}.
 \label{Chap5:Eqns:Eqn11a}
\end{equation}

By iteratively applying the above recursive relation, one obtains
\begin{equation}
 c_n \leq 4^{n-l}\left(c_l + 3 \times 2^{l-1} \right) - 3 \times 2^{n-1}.
 \label{Chap5:Eqns:Eqn11b}
\end{equation}

For the CNOT gate, applying the first optimization mentioned in Ref.~\cite{Shende2006} and taking $l=2$ gives the following inequality relation.
\begin{equation}
 c_n \leq \left( \dfrac{3(c_2 + 6) - 1}{48} \right) 4^n - \dfrac{3}{2} 2^n + \dfrac{1}{3}.
 \label{Chap5:Eqns:Eqn11c}
\end{equation}

If we consider the local equivalence classes with Cartan coordinates $(c_1, (\pi/2)-c_1, (\pi/2)-c_1)$, and $(c_1, c_1, (\pi/2) - c_1$ with $c_1 \in [\pi/4, \pi/2]$, which are conjectured to generate all two-qubit gates in two applications and contains the CNOT gate, then we can substitute $c_2 = 2$ and get an upper bound on the number gates required from this family of gates, represented by the brown coloured lines in Fig.~\ref{fig5}a, as $\dfrac{23}{48}4^n - \dfrac{3}{2} 2^n + \dfrac{1}{3}$. This is exactly one two-qubit gate less than the upper bound of CNOT gate.

Similarly, for the family of local equivalence classes with Cartan coordinates $(c_1, \pi/4, \pi/4)$, and $(\pi/4, \pi/4, (\pi/2)-c_1)$ with $c_1 \in [\pi/4, \pi/2]$, which contains the $\sqrt{\text{iSWAP}}$ equivalence class, following the arguments (up to first optimization) given in Ref.~\cite{Tang2025}, we get
\begin{equation}
 c_n \leq \left( \dfrac{3(c_3 +24)-2}{192} \right)4^n -3 \times 2^n + \dfrac{2}{3}.
 \label{Chap5:Eqns:Eqn11d}
\end{equation}

An arbitrary three-qubit gate can be implemented with at most 11 two-qubit gates~\cite{Chen2024}. However, at most 24 $\sqrt{\text{iSWAP}}$ gates are needed to implement an arbitrary three-qubit gate~\cite{Tang2025}. Hence, substituting $c_3 = 22 = (11 \times 2)$ in Eqn.~\ref{Chap5:Eqns:Eqn11d}, we can obtain the upper bound $\dfrac{136}{192}4^n - 3 \times 2^n + \dfrac{2}{3}$. This is smaller than $\dfrac{139}{192}4^n - 3 \times 2^n + \dfrac{5}{3}$, the upper bound of $\sqrt{\text{iSWAP}}$ gate~\cite{Tang2025}. 

\section{Area of the convex hull of squared eigenvalues and fractional volume of Weyl chamber covered in two applications}

\begin{figure}
\begin{tabular}{c}
\includegraphics[width=0.45\textwidth]{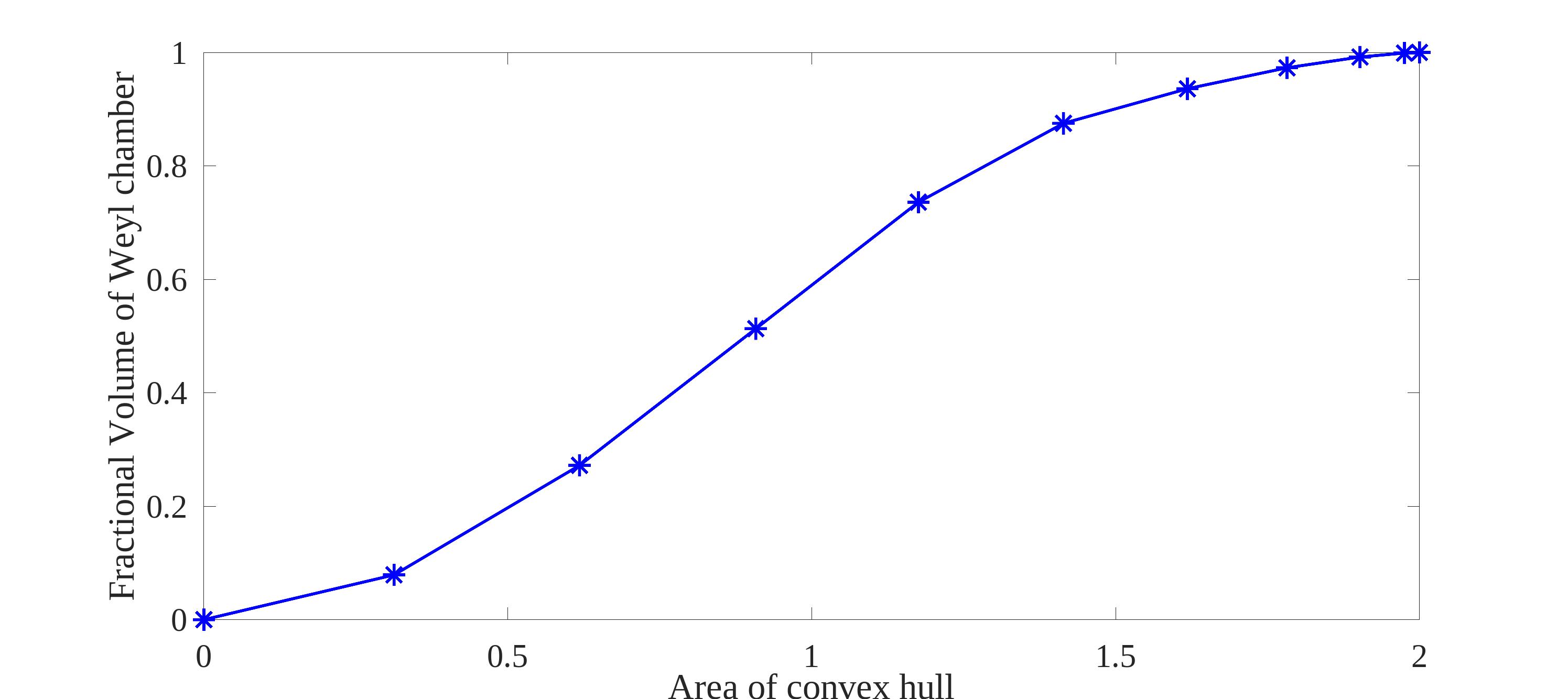} \\ (a) \\ \includegraphics[width=0.45\textwidth]{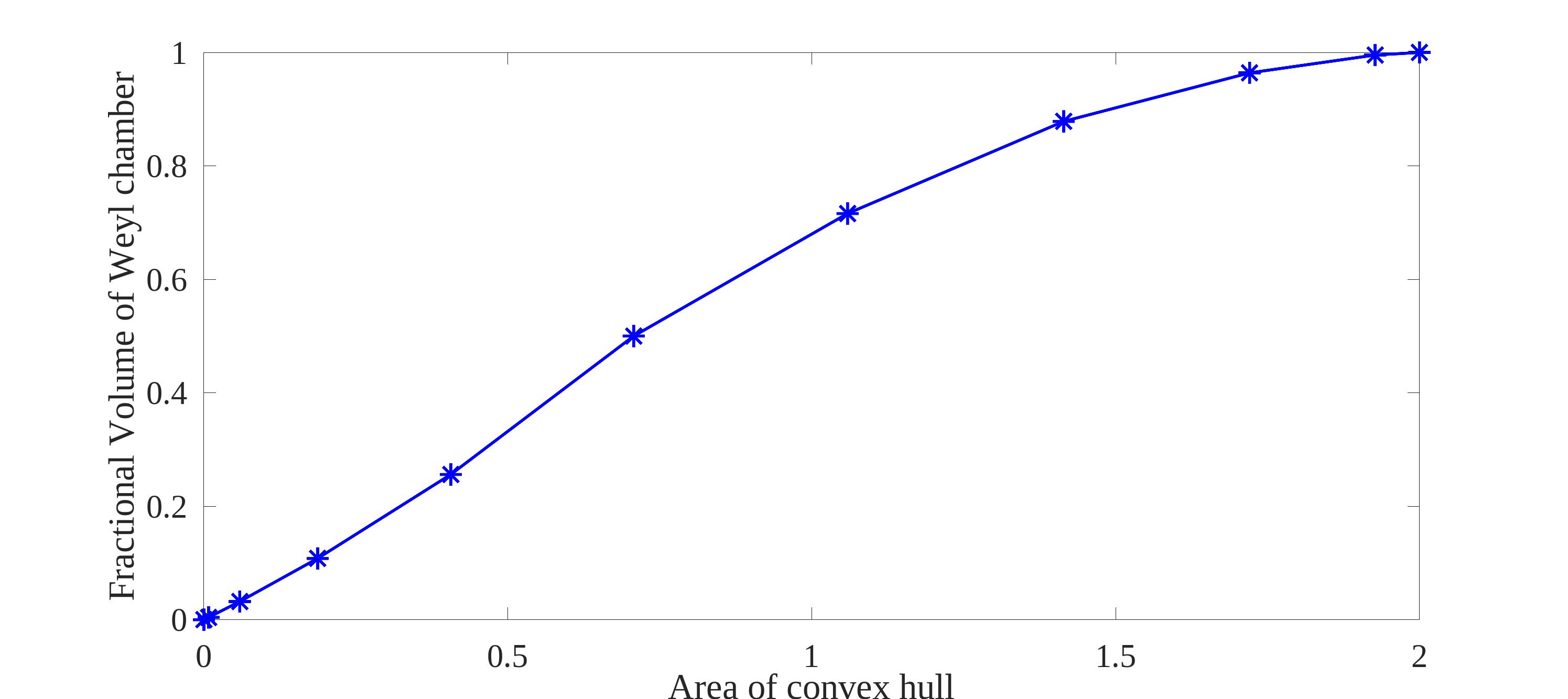} \\ (b) 
\end{tabular}
\caption{Fractional volume of Weyl chamber covered in two applications against the area of the convex hull of the squared eigenvalues of the nonlocal part of two-qubit gates with Cartan coordinates (a) $(c_1, \pi/4, (\pi/2) - c_1)$ for eleven equally spaced values of $c_1$ from $\pi/4$ to $\pi/2$, and (b) $(c_1, c_1/2, 0)$ for eleven equally spaced values of $c_1$ from $0$ to $\pi/2$.}
\label{fig8}
\end{figure}

The convex hull of the squared eigenvalues of the nonlocal part $\left( e^{ih_1}, e^{ih_2}, e^{ih_3}, e^{ih_4} \right)$ of two-qubit gates is used to provide the condition for perfect entanglers~\cite{Zhang2003}. The squared lengths of the sides and diagonals of the convex hull were shown to be related to the entangling power of the two-qubit gate~\cite{Selvan2024,Selvan2026}. In this section, we show that there exists a positive correlation between the area of the convex hull of the squared eigenvalues of the nonlocal part of a parameterized two-qubit gate and the fractional volume of the Weyl chamber covered in two applications of the parameterized two-qubit gate for the families described by the Cartan coordinates $(c_1, (\pi/4), (\pi/2)-c_1)$ with $c_1 \in [\pi/4, \pi/2]$ and $(c_1, c_1/2, 0)$ with $c_1 \in [0, \pi/2]$ (B$^\alpha$ family). For these two families, the fractional volume of the Weyl chamber covered in two applications is plotted against the area of the convex hull in FIG.~\ref{fig8} corresponding to the same values of $c_1$ used in FIG.~\ref{fig3}. For both families, the fractional volume of the Weyl chamber covered in two applications increases with the area of the convex hull of the squared eigenvalues of the nonlocal part.

It can also be verified that the fractional volume of the Weyl chamber covered by two applications of two-qubit gates with Cartan coordinates $(c_1, 0, 0)$, $(\pi/2, \pi/2, c_3)$, $(c_1, c_1, c_1)$, and $(\pi-c_1, c_1, c_1)$, which have zero-area convex hull, is zero. This suggests that there exists a positive relationship between the area of the convex hull of the squared eigenvalues of the nonlocal part of a two-qubit gate and the fractional volume of the Weyl chamber covered by two applications of the two-qubit gate. 

\section{Conclusion}

We have discussed optimal generation of two-qubit gates using the families of local equivalence classes represented by the planes that describe the mirror operation. On the $c_1 = \pi/2$ and $c_3 = 0$ planes, it is not possible to have a family without the B gate equivalence class that can optimally generate all two-qubit gates. We have conjectured that the entire $c_1 \pm c_3 = \pi/2$ and $c_2 = \pi/4$ planar regions of the Weyl chamber, excluding the B gate equivalence class, can be divided into families of local equivalence classes capable of optimally generating all two-qubit gates. It may be possible to form such families using the local equivalence classes from the other regions of the Weyl chamber; however, it should include the local equivalence classes that are invariant under inverse and the combined mirror and inverse operations, in order to generate all two-qubit gates in two applications. We have pointed out that the gates belonging to the B$^\alpha$ family and the families with Cartan coordinates $\{ (c_1, (\pi/2)-c_1, (\pi/2)-c_1), (c_1, c_1, (\pi/2) - c_1)~\text{with}~c_1 \in [\pi/4, \pi/2]\}$, and $\{(c_1, \pi/4, \pi/4), (\pi/4, \pi/4, (\pi/2)-c_1)~\text{with}~c_1 \in [\pi/4, \pi/2]\}$ are relatively easy to implement on superconducting quantum computers. We have provided upper bounds on the number of two-qubit gates required to generate an arbitrary $n$-qubit gate for two families of fsim gates that are conjectured to generate all two-qubit gates in two applications. We have also pointed out a possible existence of a positive relationship between the area of the convex hull of the squared eigenvalues of the nonlocal part of a two-qubit gate and the fractional volume of the Weyl chamber covered in two applications of the gate.


\end{document}